\documentclass[conference]{IEEEtran}
\IEEEoverridecommandlockouts
\usepackage[nocompress]{cite}
\usepackage{amsmath,amssymb,amsfonts}
\usepackage{algorithmic}
\usepackage{graphicx}
\usepackage{textcomp}
\usepackage{xcolor}
\usepackage{comment}
\def\BibTeX{{\rm B\kern-.05em{\sc i\kern-.025em b}\kern-.08em
    T\kern-.1667em\lower.7ex\hbox{E}\kern-.125emX}}
\begin{document}

\title{Digital Twins and the Future of their Use Enabling Shift Left and Shift Right Cybersecurity Operations\\
}

    \author{
        \IEEEauthorblockN{
            Ahmad Mohsin\textsuperscript{1,2}\quad
            Helge Janicke\textsuperscript{1,2}\quad
            Surya Nepal\textsuperscript{2,3}
            David Holmes\textsuperscript{1,2}
        }
        \IEEEauthorblockA{
            \textit{\textsuperscript{1}Security Research Institute (SRI)} \\
            \textit{\textsuperscript{2}Cyber Security Cooperative Research Centre (CSCRC)}\\
            \textit{\textsuperscript{3}Data61, CSIRO}\\
            \textsuperscript{1,2}Perth, WA, \textsuperscript{3}Marsfield New South Wales, Australia \\
            \{a.mohsin, h.janicke, d.holmes\}@ecu.edu.au, surya.nepal@data61.csiro.au
        }
    }

\maketitle

\begin{abstract}

Digital Twins (DTs), optimize operations and monitor performance in Smart Critical Systems (SCS) domains like smart grids and manufacturing. DT-based cybersecurity solutions are in their infancy, lacking a unified strategy to overcome challenges spanning next three to five decades. These challenges include reliable data accessibility from Cyber-Physical Systems (CPS), operating in unpredictable environments. Reliable data sources are pivotal for intelligent cybersecurity operations aided with underlying modeling capabilities across the SCS lifecycle, necessitating a DT. To address these challenges, we propose Security Digital Twins (SDTs) collecting realtime data from CPS, requiring the Shift Left and Shift Right (SLSR) design paradigm for SDT to implement both design time and runtime cybersecurity operations. Incorporating virtual CPS components (VC) in Cloud/Edge, data fusion to SDT models is enabled with high reliability, providing threat insights and enhancing cyber resilience. VC-enabled SDT ensures accurate data feeds for security monitoring for both design and runtime. This design paradigm shift propagates innovative SDT modeling and analytics for securing future critical systems. This vision paper outlines intelligent SDT design through innovative techniques, exploring hybrid intelligence with data-driven and rule-based semantic SDT models. Various operational use cases are discussed for securing smart critical systems through underlying modeling and analytics capabilities.

\end{abstract}

\section{Introduction}
Modern smart systems are composed of Operational Technology (OT), with increased integration of IT systems. These systems are collectively referred to as Smart Critical Systems (SCS), and consist of multiple Cyber-Physical Systems (CPS)\footnote{A CPS in SCSs has cyber and physical components integrated such as Supervisory Control and Data Acquisition
(SCADA), Programmable Logic Controllers (PLCs)/Remote
Terminal Unit (RTU) and Industrial Internet of Things (IIoT).}. The SCSs manage physical and operational processes across domains like smart manufacturing, autonomous production systems, and critical public infrastructures. Examples of their applications include smart grids, transportation systems, and water systems \cite{cpsics,10034656}. 
SCS facilitate realtime monitoring and control of industrial processes, ensuring safety and operational efficiency in critical environments. 
The continued convergence of IT/OT systems has significantly increased cybersecurity threats. Specifically, SCSs have emerged as primary targets for cyberattacks, capitalizing on vulnerabilities amplified by the expanded connectivity between OT/IT system components. Therefore, SCSs require 
 cyber resilient security solutions that can effectively protect these systems in the next three to five decades. 
 Recent reports \cite{ENISA2022}, indicate an escalating threat landscape for SCSs. This is attributed to unpatched vulnerabilities and unpredictable software and hardware supply chains and further compounded by inherent system complexity, insecure design, and operational silos.
 
 During a cyber attack system operators often remain unaware of these malicious activities \cite{app11219785} for a prolonged time. Earlier detection and containment, can reduce the impact and of an attack significantly. 
Cyber attacks on critical systems are often carried out by Advanced Persistent Threats (APT) groups. These APTs use 'pivoting' tactics to successfully breach networks across SCSs. For example, the 2021 Colonial Oil Pipeline ransomware attack in the US \cite{colonialattack}, the 2020 SolarWinds software supply chain attack affecting numerous businesses\cite{solarwindattack}, and the 2017 Triton malware-driven attack on a Saudi Chemical Plant \cite{Trittonattack} that disrupted safety systems and forced operations to halt, all highlight the seriousness of emerging security threats. These attack examples demonstrate that SCSs are under constant threats \cite{9763485,LI20218176} by actors who successfully target these systems and their
complex interactions and interdependencies.\par 

International security standards, such as IEC 62443 \cite{IEC62443standard} and publications from the National Institute of Standards and Technology (NIST), including the NIST Cyber Security Framework (CSF) \cite{nistframework}, emphasize the importance of integrating security throughout the entire lifecycle of systems, from design to incident detection, response, and recovery. Despite this emphasis on cybersecurity operations implementations and compliance obligations, current security tools like Intrusion Detection (ID) Systems and Security Incident and Event Management Systems (SIEMS) often fall short in capturing core assets, vulnerability scanning and identification, and correlating incidents with threats and vulnerabilities. Consequently, these tools struggle to safeguard critical systems effectively or respond to cyber incidents promptly. While their enabling technologies such as Artificial Intelligence (AI) or knowledge-based systems alone are not capable enough to manage cybersecurity operations effectively. 

Digital Twins (DTs), are promising technological solutions which can effectively tackle the present and anticipated cybersecurity challenges over the next three to five decades of future critical systems. DTs hold the potential to risk mitigation, providing platforms for security evaluation through simulations and testing the efficiency of cybersecurity measures, and even predicting realtime security threats. 
\begin{figure}
\centering
\includegraphics[scale=0.44]{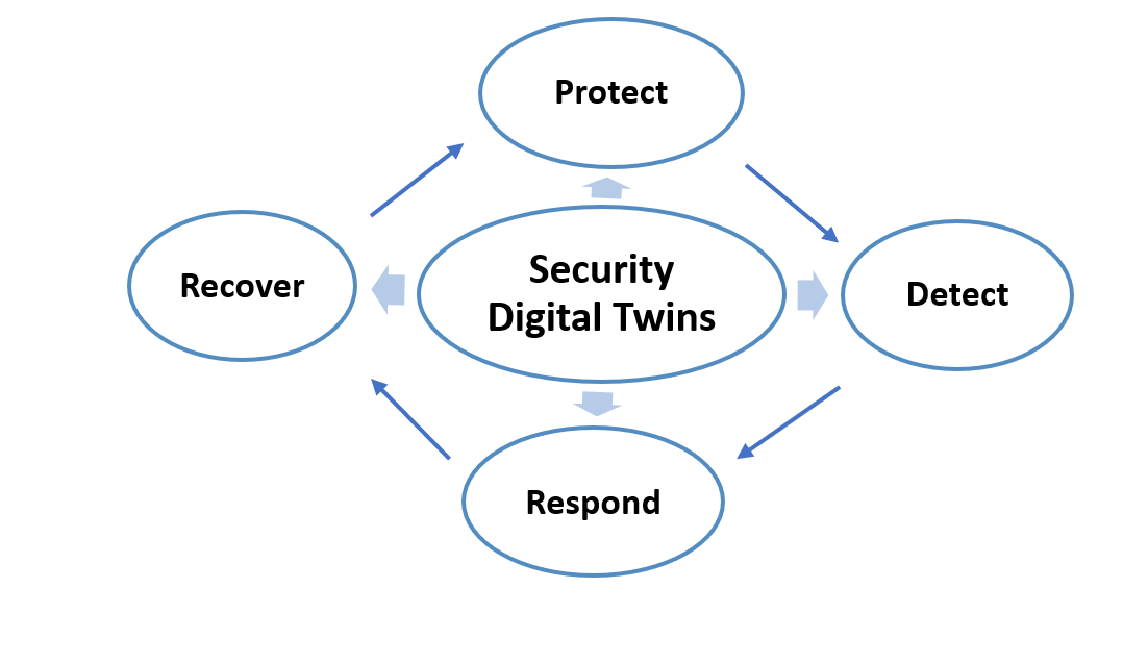}
\caption{Security Digital Twins based cybersecurity operations}
\label{fig-1}
\end{figure}
A DT is a virtual counterpart of a physical system that mirrors their behaviors, synchronized with data connections, enabling dynamic analysis, predictive insights, and informed decision-making for enhanced operational efficiency and innovation across industries \cite{DTclassic}. A DT monitors the physical processes, environmental, and operational parameters of a SCS with simulations allowing optimization and predictive maintenance analytics \cite{schluse2016simulation,pires2019digital}. 
DTs are commonly developed using rule-based, semantic annotation, and Machine Learning (ML)/AI-based approaches to facilitate advanced analysis and simulations.  We propose that just as DTs are effectively employed across Industry 4.0, spanning tasks like asset monitoring and product lifecycle management, involving automated and human-assisted interventions, a similar cybersecurity operations strategy can be adapted for SCSs \cite{holmes2021digital}.

The main thrust here is to improve the cybersecurity operations of SCSs employing Shift Left and Shift Right (SLSR) design paradigm at both design and runtime by developing their security-driven DTs to cope with system dynamics and complexities while improving cybersecurity operations enhancing system resilience and trustworthiness at scale. The Security Digital Twin (SDT) is a software-driven solution focused on solving SCSs cybersecurity tasks. A SDT can ensure the separation of concerns by ensuring cybersecurity operations are at first initialized and fixed in a virtual replica and then later can be reflected in the SCS without interrupting core business and safety operations. A perspective SDT with enabled cybersecurity operations following NIST CSF is visualized in Figure \ref{fig-1}. \par 
Considering the attack on Saudi Chemical Plant \cite{Trittonattack} in which attackers targeted plant safety systems by manipulating OT systems through advanced malware, an SDT could be used to emulate and monitor plant safety operations states, sending them realtime instructions to maintain required safety measures if certain events occurred in DT models of the plant. An SDT could also help reveal an attacker's presence in the network early during the cyber breach, which might otherwise remain undetected for many months.\par 

The SDT ensures reliable and synchronized data flows from the CPS to orchestrate behaviors for effective security modeling and analysis. Therefore, physical control components such as PLCs and IIoTs should have more reliable connections and data streams for SDTs to model and simulate respective behaviors in a DT environment to orchestrate cybersecurity operations. To overcome these challenges, we explore the introduction of Virtual Components (VC) replacing components of the CPS with virtualised counterparts\footnote{Throughout this paper, a virtual CPS component (VC) represents different critical components such as PLC/RTU and IIoTs} directly manipulating and controlling physical processes from Edge or Cloud environments and feed reliable data to  SDTs enabling their seamless integration, design, deployment, and testing for intelligent cybersecurity operations. The VCs in the SDT as building blocks then support the cybersecurity of SCSs throughout the lifecycle employing the Shift Left and Shift Right (SLSR) design paradigm. The SLSR design paradigm emphasizes the importance of security starting from the very beginning of system design (Shift Left) and continuing as the systems operate (Shift Right) using SDT. This ensures  SCSs are more resilient and dependable against emerging cyber threats.  

We propose innovative SDT-driven cybersecurity operations of the future SCSs with a new design paradigm of SLSR employing VCs. With this architectural shift in design, we explore and discuss the enhancement of SDT modeling capabilities incorporating a mix of state based modeling, semantic technologies, Knowledge Graphs (KGs), and emerging ML and generative AI approaches. We also identify research challenges that need to be overcome for the implementation of DT-driven cybersecurity of future systems.

\par 
The paper is organized as follows: Section II introduces the concept of virtual twins using SDTs and the SLSR design paradigm. In Section III, core aspects of security-driven DTs are presented as SDTs for vital cybersecurity operations. Section IV explores emerging modeling and simulation methods for building advanced analytics capabilities for cybersecurity using SDTs. Section V covers related techniques, and Section VI outlines research challenges. Finally, Section VII concludes with future research directions.

\begin{figure*} 
\centering
\includegraphics[scale=0.35]{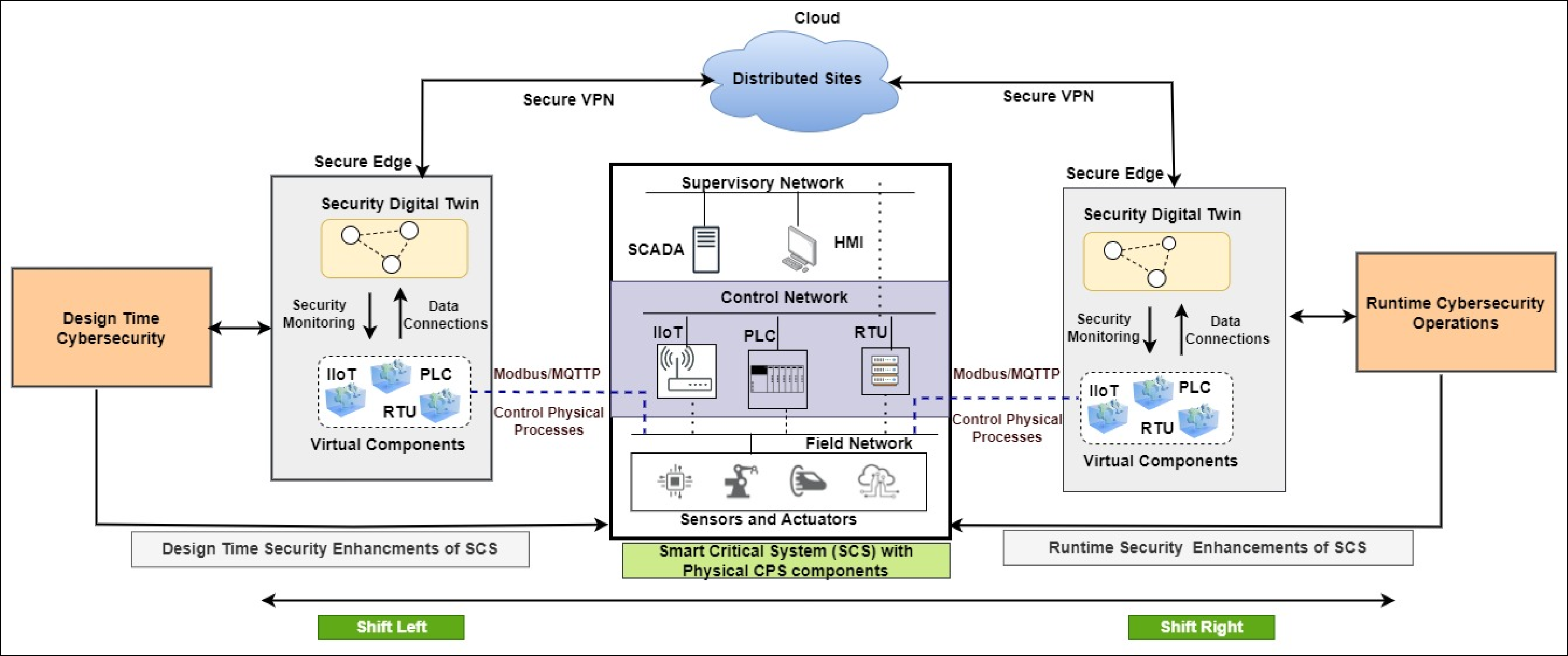}
\caption{Design Paradigm Shift: Virtual Components based on Security Digital Twins  enabling cybersecurity operations with Shift Left and Shift Right}
\label{fig-2}
\end{figure*}

\section{Shift Left and Shift Right Design with Virtual Components}
This section introduces the concept of SLSR design paradigm for SDTs using VCs as building blocks across the SCS security lifecycle. We first establish the rationale for VCs replacing components of the CPS, and then describe how the SLSR design paradigm works with SDTs. 
\subsection{Virtual Components}
\textbf{Motivation:}
The existing class of SCSs consists of a variety of diverse CPS hardware and software components. These components function across different levels following Purdue model \cite{Purduemodel} and as such consistent and reliable interaction of these components as CPS to DTs is always at risk for monitoring of core system properties during operations. For example, consider a scenario involving a smart grid, where CPS components such as PLCs play a pivotal role in overseeing and controlling sub-stations. The primary responsibility of these PLCs is the continuous monitoring and control of critical system processes like voltage control, and power flows, 
while communicating critical systems information and other related factors where they work in a closed loop with DTs \cite{smartgrid2, smartgrid1}. 

When addressing the implementation of DTs in SCS, security needs to be considered as the key driver to ensure system resilience. DTs must have secure and consistent connectivity to the CPS for effective monitoring and intelligence gathering. Establishing and maintaining an uninterrupted connection and synchronization of data between the physical and digital counterparts can be a significant challenge. The criticality of meaningful data ingestion by the DT is vital for its operational resilience.
 These challenges persist from the early stages of design, development, testing, and security analysis through to deployment and operationalization. 
With the advent of increased WiFi performance connectivity using 5G/6G, available protocols\footnote{Communication and Security protocols in use between control devices and field devices i.e., (OPCU UA, Modbus) for PLCs/RTUs and (HTTPS/MQTT) for IIoTs when interacting with sensors and actuators} \cite{PIVOTO2021176}, the Physical  CPS components can be replaced with a Virtual Component of similar functionalities and has the ability to integrate with SDTs and SCSs. These VCs hosted in edge or cloud computing environments reside closely with SDTs and send operational data to DTs based upon reference architectures such as the Purdue model \cite{Purduemodel}. A VC is a programmed software component to replicate the logic and behavior of CPS and is integrated with physical components (sensors and actuators) at the device level to directly interact and control physical processes, refer to Figure \ref{fig-2}. The introduction of VCs into SCSs such as smart grids from constrained sensors and actuators, enable data capture at the Edge environment \cite{9211890, virtualisationIIOts} address interoperability issues for data exchange between the SCS and SDT, delivering a more structured data framework 
providing advanced analytics for improved security operations. 

With VCs, the likelihood of data loss for SDT is minimized, and a more reliable source of truth is enabled with increased availability of data to an SDT. The scalability of security operations using VCs is more flexible and easy to manage as additional components can be added to enrich SDT features to efficiently monitor and support SCS functionality.  
A VC supporting interactions with these devices and software components can gather contextual information about vulnerabilities and threats, which can be fed into the SDT for improved security operations, otherwise not available for collection from the traditional CPS. 

\textbf{Advantages:} 
(1) More reliable data integration to SDTs enables constant information feeds to build SDT behaviors with the ability to store historical and realtime data in the edge and cloud platforms enhancing the scalability of SDTs for application in various security use cases which otherwise would not have been possible. 

(2) The SDT can seamlessly intervene VCs when certain changes or conditions for normal operations are violated and indicated in SDTs. Respective business safety and security operations can be managed by re-adjusting VC parameters and program logic with minimal impacts. This reduces overall downtime in case of anomalous behaviors in field devices and physical equipment failures during operations.

(3) Remote security operations with VCs can be achieved where each VC receives alerts and recommendations from an SDT regarding security incidents with an intervention functionality to improve upon security controls through implementation of identity and access management, minimizing security threats, and improving incident response reaction  time as compared to a SCS where cyber incursions may remain undetected for months.

\subsection{Shift Left and Shift Right Design Paradigm}
Figure \ref{fig-2} outlines the SLSR paradigm for cybersecurity operations in SDTs. Shift left design based on SDT ensures the integration of security design tasks at design time ensuring early SCSs 
security testing, simulation, identification and mitigation of vulnerabilities. In rare cases, the shift left design approach is used for evaluating functional correctness and performing certain simulations and validation using DTs, while the shift left design for cybersecurity brings together new opportunities for security teams to evaluate system security aspects using SDTs early at design time. Shift left approach only suffices design time security using SDTs while SCSs need to be secured, monitored, and analyzed for their security throughout the lifecycle after they go into the production phase. To this end, we propose the use of shift right ensuring runtime/operational cybersecurity of SCSs employing SDT modeling capabilities with various operational cybersecurity use cases. Security testing, monitoring, simulation, and analysis are considered for shift right SDT-based cybersecurity of critical systems. 
\par 
We discuss the potential of SDT-supported, shift left and shift right cybersecurity capabilities in the next section.     
\section{Digital Twins for Cybersecurity Operations}
Cybersecurity operations for SCSs require the presentation of diverse perspectives from overall system security. This facilitates evaluating and preparing for potential threats, vulnerabilities, and attacks, particularly when these systems operate in complex environments. Current SIEMs, together with endpoint detection and IDS tools, offer either limited intrusion detection/prevention capabilities or a very basic ability to monitor specific SCS components in networks. Moreover, these security tools and other state-of-the-art security modeling solutions are not able to augment and analyze for timely incident response in case of security incidents.  An SDT capability offers many opportunities and features for security teams and researchers to mitigate security risks early in the security design phase while ensuring resilience to cyber attacks at runtime. \par
\begin{figure*} 
\centering
\includegraphics[scale=0.32]{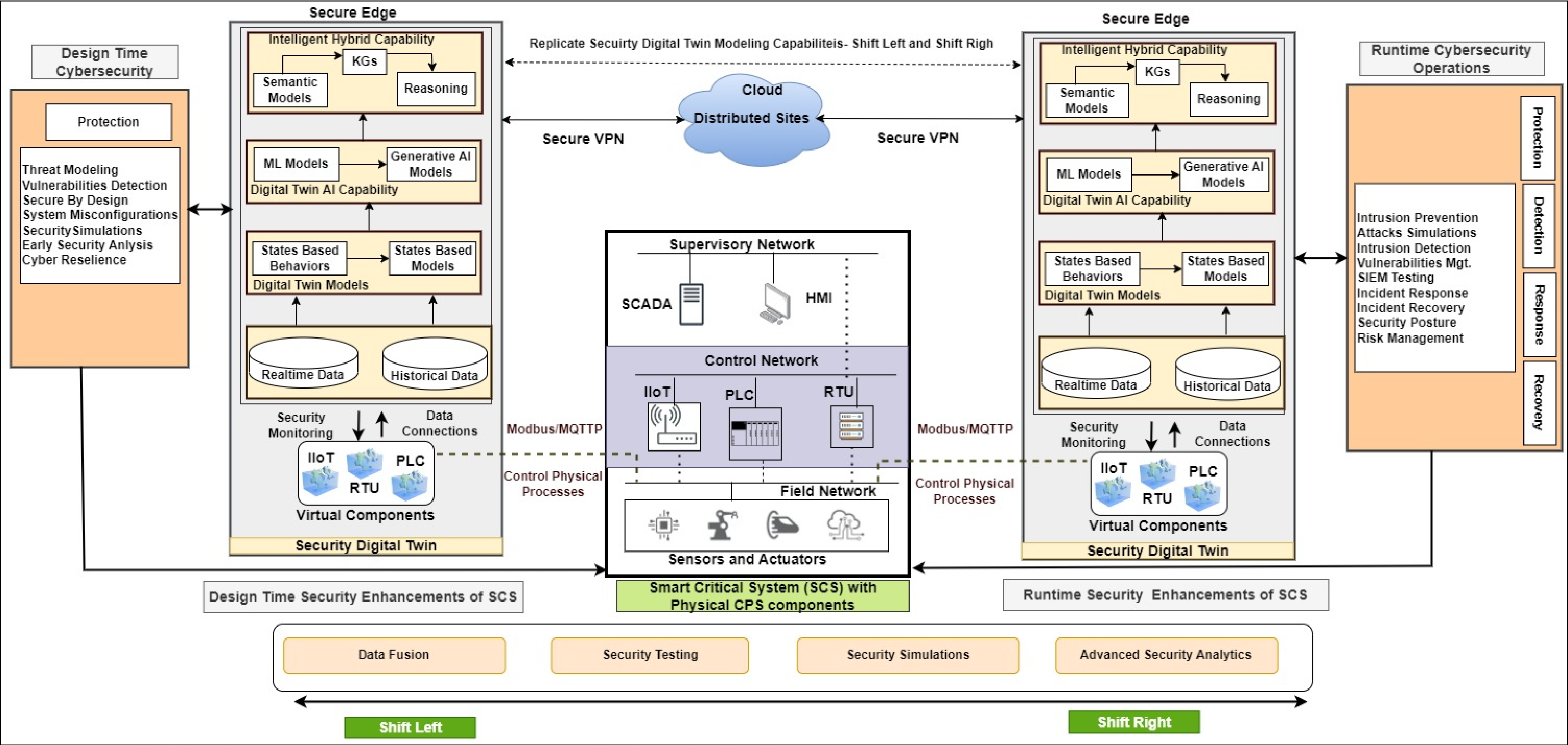}
\caption{Security Digital Twins: Cybersecurity Design time and Runtime Modeling Capabilities with Shift Left and Shift Right (SLSR) for Smart Critical Systems. Virtual components work closely with SDTs while providing realtime and historical data about SCSs CPS components. Modeling capabilities for SDTs can be replicated from design time to runtime for improving the cybersecurity of SCSs}
\label{fig-3}
\end{figure*}

 
In an SDT, the fusion of historical data with realtime data offers various multi-dimensional features such as simulation/testing of critical assets, various types of security analytics, security controls optimization, and interventions, depending upon security incident severity and context. Figure \ref{fig-3} depicts VCs integrated with field devices (sensors and actuators) to enable the availability of different types of data states for security-driven DTs. These are used to design intelligent and robust security models supporting broad security use cases by delivering protection, detection, response, and recovery at both the design and runtime of SCSs.\par 
As visualized in Figure \ref{fig-3}, the VC obtains the physical process operational information through sensor outputs from field devices and feeds this data to the SDTs for cybersecurity operations monitoring. The current process state of 
various physical processes is captured through VCs, which then defines their functional behavior as a collection of states over a period of time within SDTs. The data and process states contain information about critical systems, their sub-systems, and interactions (what type of sensor data VCs read and what inputs they send to actuators as well as SCADA and HMIs) to carry out physical/business processes. The data can be specified in XML/JSON formats such that it describes the CPS business operations conditions, process controls logic, and threshold values for certain operations, such as voltage values and power flow in a smart grid for a power generation device should not cross certain threshold values to ensure safe operations. All this information from VCs and their physical counterparts is used to enrich SDT models. The SDT models are further leveraged with ML and Large Language Models (LLMs) capabilities to provide various cybersecurity operations use cases. Similarly, semantic models, knowledge representation, and reasoning capabilities enable hybrid intelligence, see Figure \ref{fig-3} for SLSR SDT-based modeling capabilities.\par
In the following, we discuss various cybersecurity SLSR operations using SDTs-driven capabilities. 
\subsection{Security Digital Twins-Enabled Systems Security with Shift Left and Shift Right Design Paradigm}
\subsubsection{Protection}
The protection of SCSs is vital to safeguard against rising threats and attacks. The protection use cases enabled through SDTs are described in detail aligned with SLSR paradigm \par 
\textbf{Addressing Security Misconfigurations:} 
The SCS, with diverse components and interconnections, face many security misconfigurations such as software, hardware and network-related misconfigurations. Since SDTs are virtual replicas of exact system components, they can easily simulate and mimic the component's configuration details. For example, a virtual component of a CPS with SDT can provide network connection details, and communication protocols, and present the way internal components, such as virtual PLC perform interactions with Sensors and actuators. This way SDT can simulate network configurations that are specified and compared with baseline configurations. If network or hardware/software configurations have gone through certain changes, then these can be detected in SDT simulated environment to identify misconfigurations and fix those providing security teams with the opportunity to address these issues early in the lifecycle \cite{DTSOAR}.  Within the SDT environment,  CPS configurations vital to system security can be tested by generating multiple security scenarios verifying and evaluating each configuration against certain attacks and exploitable vulnerabilities both at design and runtime to further enhance the security posture of systems. \par 
\textbf{Threat Modeling:} 
Threat modeling of SCSs is a vital cybersecurity feature for SDT. An SDT can model and simulate internal and external interfaces of CPS through VCs for malicious actor's entry points to the system, helping to identify system vulnerabilities and associated risks, especially at design time. 
This way threat modeling can help explore various interactions between interdependent system components of complex SCSs at early stages of development. For example, threat modeling techniques such as STRIDE and DREAD \cite{8260283} can be applied to identify vulnerabilities at each interface both internal (between VCs to field devices and HMIs/SCADA systems) and external (OT systems providing access to third-party users and applications). The vulnerabilities can then be analyzed and mapped to CVE\footnote{Common Vulnerabilities and Exposures, https://cve.mitre.org/} for multi-stage attacks on SCS where each CVE can be assigned to different attacks. Take the example of PLC related vulnerabilities found in program logic which can be mapped to CVEs and CWE\footnote{Common Weakness Enumeration, https://cwe.mitre.org/} entries for risks rating and their severity levels. 
The analytics capability in a SDT employing ML \cite{Kaur2020}  in a simulation environment can help visualize asset threats and their interdependencies for addressing associated risks at various stages of critical systems. 
Both at design and runtime SDTs can leverage semantic models and KGs for effectively building threat models showing relationships about potential threats, vulnerabilities, and their origins.  \par   
\textbf{Vulnerabilities Fixing:} At Design time SDTs can help fix and patch vulnerabilities through threat modeling and security testing of SCS components early in the lifecycle. This helps security teams to minimize potential threats with less number of vulnerabilities when the system eventually goes into the operational phase.  

\textbf{Intrusion Prevention:} 
With the shift right perspective for operational security, SCSs can be protected against attacks at runtime by employing SDTs. The existing DTs with behavior-specific security patterns using data-driven and semantic technologies \cite{9453553,info14020095} provide certain ID capabilities; however, ID alone is not sufficient for the protection of CPS and related components with increased convergence of IT/OT systems. In relation to SDT right shift approach, Intrusion Prevention (IP) models must be designed to observe system dynamic states capturing system physical properties such as CPS sate-space-based methods for behavior change prediction. With these predictions, the SDT can determine if particular physical process-related sensor values violate threshold values and reach unsafe or undesired states\cite{9527938}. With the advanced prediction of unwanted behaviors, certain security incidents can be avoided. The SDT can identify relevant assets from these predictions to activate security countermeasures, such as improving security controls around PLCs/HMIs for unauthorized access to safeguard against false data injection and Man-in-the-middle types of attacks. 
\par 
\subsubsection{Detection}
Intrusion and anomaly detection using SDTs show promise for accurate and reliable detection of security anomalies across physical and virtual components of critical systems \cite{9453553}.\par 
\textbf{Intrusion Detection:} By utilizing realtime and historical data repositories from physical and virtual components, intelligent solutions are created to enhance the features of ID system using SDTs, enabling malware and unauthorized access detection at different levels of the system. For instance, specific unusual patterns, such as changes in system configurations in SDT, can be observed using knowledge-based detection with static data (historical data) \cite{IDS-staic-DT} from sensor devices, system logs, and network traffic. The SCS  behavior models within SDT can be updated with ongoing changes in SCSs complex environment to simulate malicious activities and perspective attacks. Similarly, certain rules can be applied to oversee the behavior of Virtual Components for security and safety incidents. By employing ML techniques \cite{9438560, Varghese_2022}, dynamic ID can be facilitated using both historical and realtime data by behavior refinement. This allows for continuous monitoring of unusual activities from DTs to CPS.  

\textbf{Digital Twins-based Security Tools:} 
Shift right security operations features from SIEM and Extended Detection and Response (XDR) tools can be simulated and tested for SCSs before such systems are actually implemented to assess their suitability for various security use cases. Deployment and testing SIEM tools directly on critical systems are quite expensive and cumbersome, especially testing their correctness of incidents detection, alert generation and response. SIEM and XDR systems can gradually be tested with critical CPS components within SDTs with complete life cycles of events indigestion, pre-processing to detection, and mitigation testing in SDTs \cite{DTSOC} for better configuration and performance. Selected security functions with SDT against selected CPS components are validated and passed through scenarios, then the gradual deployment of SIEM/XDR can be carried out across real-world SCSs. This enables the correction of security functions' applicability and avoids security failures in production.

\subsubsection{Response and Recovery}
Business and safety-related operations delayed response and recovery seriously effects an organization's reputation and financial revenues in the eve of cybersecurity incidents. From an SCS perspective, response and recovery to cyberattacks are not straightforward. Take the example of the Solarwinds attack, which is believed to have started in 2020 \cite{solarwindattack}, which was discovered in late December 2020, and initial response and recovery procedures started in mid-2021. Security Orchestration, Automation, and Response (SOAR) aided with ML and rule-based advanced analytics are used for incident response activities. However, SOAR applicability for critical systems has challenges with additional OT components, IT/Enterprise systems, impacted stakeholders, and business function differences and operational silos, adding technical complexities leading to delays in security incidents response and recovery. One way to overcome these challenges is to test and evaluate incident response and recovery in actual CPS environments across OT/IT landscape, which is usually very expensive and nearly impossible. SDTs allow security teams and security analysts to test/simulate various response scenarios\cite{info14020095}. The SDTs with enough data from CPS can be used as effective incident response training platforms for critical systems. A digital twin-based SOAR for critical systems provides innovative ways to respond and recover from advanced and complex attacks\cite{DTSOAR}. 
\par 
\textbf{Response Scenarios Simulations:}
Cyber attack incident response scenarios can be simulated within SDTs, before implementing them in the real world. Penetration testing of CPS is conducted in SDTs to evaluate the impact of various attacks on actual systems and their tailored responses. During the attack, the impacted IT/OT system components can be isolated or quarantined to verify the spread of malicious code or artifacts within the SDT. Security teams can effectively test various containment, remediation and mitigation strategies within SDT before these strategies in case of such incidents can be implemented in actual critical systems environments. During response simulations, discovered security vulnerabilities can be prioritized and fixed through digital twin automated recommendations early in the lifecycle.

\section{Towards Intelligent and Reliable Security Digital Twins}
The SDT driven cybersecurity operations elaborated in the above section use a variety of modeling techniques to bring intelligent, security analysis capabilities for smart critical systems. 
 The modeling capabilities of SDTs mostly rely on rule-based, physics-based, and machine-learning methodologies \cite{10049398, Kaur2020, austin2020architecting}. However, these methodologies are often explored and implemented independently within DTs without cohesive support for various cybersecurity operations in the context of SLSR design. Consequently, the security use cases commonly lack centralized analytics employing integration and automation of SDTs, which is essential for providing comprehensive support to cybersecurity operations.
 Security researchers and practitioners should adopt innovative approaches for designing the architecture of physical/Digital Twins, enhancing the security intelligence capabilities of SDTs, and enabling dependable and seamless automated support for decision-making throughout the systems lifecycle. \par 
\textbf{Integrated and Automated Cybersecurity:}
An integrated and automated approach leveraging intelligent modeling paradigms such as AI/ML, rule-based, knowledge-based, and ontological approaches can be combined to enable innovative cybersecurity operations use cases within SDTs. Following such an approach integrated security modeling within SDTs can be proven useful where each SDT model can use a variety of CPS-generated data to augment cybersecurity operations. As visualized in Figure \ref{fig-3}, realtime and historical data are fed to digital wins for developing SDT models. The data types ingested through physical and VCs can be further categorized into static descriptive data, dynamic 
critical assets data, dynamic operational environments data, and semantic data. Each of these data categories is aligned with SDT modeling capability for cybersecurity operations. These modeling capabilities can be categorized into state-based behavioral security models, ML-based Security models, and knowledge-based hybrid security models. These security models potentially add advanced analytics, simulation, and descriptive security capabilities for cybersecurity operations within SDTs. 

In the sections below, we explore and discuss the development and enrichment methods of these modeling capabilities for digital twin-driven cybersecurity operations. 
\subsection{AI Capability for Security Digital Twins}
Digital Twins based on AI/ML modeling techniques are suitable to leverage big data for advanced analytics, operational intelligence, and optimization.
Enhancing SDT-driven cybersecurity operations involves leveraging various ML/DL algorithms and design strategies like anomaly detection, security prediction modeling, and attack forecasting with state-based and historical data \cite{MLcybersec}. While several research endeavors have focused on designing DTs for specific security tasks, many techniques primarily rely on static data states, using historical data to train ML models for identifying security anomalies and intrusion detection. Instead of these techniques, we require a bottom-up approach. Here, ML/DL models for SDTs receive realtime and historical data from SCSs, which helps establish baseline behaviors for CPS. During model training, this data is correlated to detect outliers and predict potential threats and malicious activities, see details in Figure \ref{fig-3}. For critical systems, SDTs can collect large amounts of data from CPSs for training AI/ML models which can be tailored to particular cybersecurity operations by combining classification algorithms i.e., Support Vector Machine (SVM), Artificial Neural Network (ANN), Generative Adversarial Networks (GANs) and Gradient Boost (GB) for intrusion and anomaly detection related cybersecurity tasks \cite{9438560,Varghese_2022}. \par 
The underlying SDTs models can take advantage of AI models to improve security risk management through threat modeling, incident response management, and similar use cases implementing the SLSR approach. 

\subsection{Generative AI-based Security Digital Twins}
\textbf{Attacks Simulations through Generative AI:}
Generative AI-based attack scenario generation using GANS for testing and evaluation of CPS security can assist in determining their ability to resist sophisticated attacks and improve the overall cybersecurity posture. An SDT using Generative AI can generate attack simulations representing real-world attackers' tactics. For example, a simulation of a cyber attack targeting CPS components such as HMI/Engineering Work Station can reveal the tactics, techniques, and procedures used to breach OT systems and can assist security teams in proactively formulating an effective response strategy to mitigate cybersecurity threats and incidents. 
As depicted in Figure \ref{fig-3}, Large Language Models (LLM) are an emerging type of generative AI \cite{kucharavy2023fundamentals} used in combination with traditional ML/DL models to evaluate and support security operations using DT technology. The malicious code and its infiltration within the CPS support meaningful impact analysis of AI-generated code exploiting system vulnerabilities. As a result, gaps/weaknesses in
current security controls of SCSs, policies and vulnerabilities can be preemptively addressed through the utilisation of the SDT.\par   

\textbf{Generative AI for Incident Response and Recovery:}
By employing LLM-based pre-trained models such as Open AI ChatGPT \cite{openai2023} and Microsoft Security co-pilot fine-tuned \footnote{ https://www.bleepingcomputer.com/news/microsoft/microsoft-brings-gpt-4-powered-security-copilot-to-incident-response/} models into the SDT landscape, security researchers can leverage the intelligence and agility of LLM agents for cybersecurity operations augmentation. This enhances security team agility by simplifying security operations through LLM SDT agents to interpret and relate to the context of security incidents, including VCs and the physical environment.
LLM-based simulated behaviors of SCSs are supported by large dataset feeds of realtime and historical observations from physical and virtual counterparts see Figure \ref{fig-3}.

Furthermore, these LLM-based SDT agent datasets seamlessly incorporate security appliance data sourced from CPS, encompassing data from firewalls, network logs, SIEM, and EDR logs. This holistic approach provides a comprehensive insight into system architecture and the flow of information traffic, empowering incident response efforts. At scale, security teams can effectively pinpoint and comprehend potential threats by employing natural language queries to inquire LLMs about security vulnerabilities, threats, and the overall security stance of critical systems. This, in turn, equips them with the knowledge needed to proactively prepare and respond to potential attacks. SDTs, powered by LLM security agents, enable concurrent investigation of ongoing attacks for quicker containment, remediation, and mitigation, reducing the mean time to protect, detect, respond, and recover critical systems.

\subsection{Semantic Technologies Capability for Security Digital Twins}
Semantically enriched ontologies and KGs \cite{semantickGsDT} can be leveraged to create an intelligent and context-aware SDT modeling capability through the integration of metadata feeds from the CPS devices, sensors, and actuators. Semantically enriched SDT modeling capability is used to analyze and simulate the behavior of the system under different operational circumstances and facilitate meaningful predictions for critical system security concerns.
\par
\textbf{Knowledge Representation and Security Modeling:} From a security Digital Twins perspective, ontological reasoning and KGs can further facilitate the execution of complex security analysis using entity resolution, clustering, and classification.
Machine reasoning in support of cybersecurity operations can be implemented using rulesets and inference techniques. New threats are identified from the existing knowledge base, which can further be used for in-progress attack detection and a greater understanding of attack paths and pivot points. For security ontologies development, ontology languages such as Resource Description Framework (RDF) and (Web Ontology Language (OWL) are commonly used \cite{semantickGsDT}. The SDTs enrich the contextual data of CPS with underlying ontologies attributes against threats and vulnerabilities for developing situation aware cybersecurity operations. Data models with different types from CPS, including physical and virtual assets, network logs, and incident reports, can be aggregated to develop multiple security ontologies, each with a specific objective for cybersecurity operations \cite{austin2020architecting}.  \par  
\textbf{Knowledge Graphs for Cybersecurity Operations:}  Constructing KGs for SDTs in the context of cybersecurity operations involves combining metadata from various sources to create a structured representation of the relationships, entities, and attributes within the DT environment. Incident detection can be intelligently done through KGs capabilities. Detected cyber incidents within SDTs can be co-relations to inference and track the origin of attacks and vulnerabilities discovery and their paths with inner-related components using KGs data models. The KGs enriched with CPS states and asset specifications data with underlying ontologies can be used for incident response and management. In this regard, KG-based annotations aided with MITRE ATT\&CK framework provide valuable insights into cyber threat techniques and respective Indicators of Compromise (IoC) for IT/OT systems attacks mapping\footnote{https://attack.mitre.org/}. This mapping can be aligned to attack vectors for building automated attack paths to enable automated IoCs identification against each attack tactic employing Semantic modeling and KGs within SDT capability. These MITRE ATT\&CK-based models can then be used in incident playbooks for devising better response and mitigation strategies.  

\section{Related Techniques}
In this section, we evaluate the commonalities and differences of techniques with their pros and cons for the cybersecurity operations of critical systems in relation to digital twins. \par 
Closely related to SDTs, hardware and software-based emulators and simulators are used for CPS testing and validating cybersecurity operations \cite{mcdonald2010modeling}. 
However, cybersecurity modeling and simulation methods similar to this do not have the capability to monitor system security operations with realtime synchronization.
The SDT with shift left and  shift right security testing using intelligent modeling capabilities, are both flexible in implementation and effective in operation, remaining consistent throughout the system lifecycle. 

Cyber ranges are useful platforms for training, identifying system vulnerabilities, associated threats, and attacker's tactics for exploitation of vulnerabilities using tailored scenarios \cite{YAMIN2020101636,8842802}. 
Security defenders can test and validate security hardening techniques by utilizing different security appliances, including EDRs, SIEMs, and firewalls designed to detect and protect against bespoke attacks. In comparison to SDTs-based security operations, cyber ranges operate mostly with hybrid system components, only offering security testing and training capabilities, which are mostly manually simulated. In contrast, SDT technology offers such automated features as security-driven behavior state modeling, simulation, replication and advanced support diagnostic analytics. Security system design, prototyping, deployment, commissioning and standards compliance phases are each the recipient of increased capabilities when using SDTs. The use of SDT technology enhances routine security administrative tasks when performing system security testing and analysis activities.

\par 
A number of AI/ML modeling approaches \cite{MLcybersec,9438560,Varghese_2022} have been developed as standalone prototypes as well as integrated into existing security appliances. As discussed in Section IV, these techniques have a certain level of intelligence for security incident prediction and detection. However, unlike an SDT, they do not possess the essential contextual knowledge and situational awareness necessary 
for the meaningful visualization of cybersecurity incidents, the tracking of incursion sources, and the interaction activity between system critical components. 
Moreover, standalone AI/ML-based security tools and appliances are not integrated in a way to streamline security operations with design and operational scenarios to test and evaluate the cybersecurity of SCSs.\par 
The SDTs do not necessarily replace existing security appliances and tools, but they do enhance their existing ability to understand the complexities of SCSs with state-based and semantically enriched KGs improving understanding the context of security incidents through simulated analytical techniques which conventional security modeling tools alone cannot achieve. 


\section{Research challenges}\label{AA}
While Digital Twins show promise for driving cybersecurity operations, there are challenges in their design, architecture, and enabling technologies. These issues require attention from security researchers and practitioners. In the following, we highlight some of the challenges that demand further exploration and research endeavors from academic and industry practitioners:
\par 
\textbf{Virtual Components Security:} The addition of VCs 
suggests many significant security benefits but also introduces associated risks and security architecture challenges. The perspective VCs, embedded in either Edge or Cloud environments, may expose particular security weaknesses and provide entry points for attackers to gain access to the SCSs if not designed securely. 
In this regard, security architecture patterns, both for Edge and cloud-hosted DTs should be designed to cater for security controls incorporation at various levels\cite{virtualisationIIOts}. Specialized security verification and validation methods should be used to evaluate the security of VCs and their environments. Data transfers from field devices to Edge, pre-processing, and further distribution to cloud and SDTs require the exploration of new data encryption methods across the channels while the shared responsibility models between edge and cloud should be analyzed and understood by stakeholders. \par   
\textbf{Digital Twins Security:} The SLSR integration of SDTs into SCSs introduces certain risks as the left and right interactions increase the opportunities for adversaries.
One of the heightened security risks that an SDT introduces to the CPS is that of additional attack vector vulnerabilities. 
The result of its digitalization of the physical hardware, the SDT digital hardware, and software components adds to the security exploitation opportunities for malicious actors. 
For example, by unauthorized access through the SDT of a smart grid, an attacker can gain full visibility of core grid functions and then pivot into access to critical assets. Exploiting security weaknesses in SDT, a malicious actor could potentially launch advanced attacks on the smart grid leading to the disruption of essential services, and compromising the safety and security of the system. 
To address these issues, security researchers and system architects should review methods of system security and modeling integration with a view to the development of secure cybersecurity frameworks with the implementation of multi-layer protocols prioritizing cybersecurity\cite{holmes2021digital}. Emerging technologies such as Blockchains \cite{blockchainSDT} can be employed for SDT identity and access management across SLSR interfaces to overcome such issues.\par
\textbf{Interoperability and Communication Issues:} The interacting twins, i.e. SDTs and CPS physical counterparts, use different data formats, and protocols, for communication. Integration of both twins is a complex and challenging task due to the non-existence of standardized interfaces and governance. This adds to performance overheads and delays in data processing and system response times, potentially affecting control and decision making in CPS via SDTs.
To address these challenges researchers should focus on developing standardized communication protocols and data representation formats, as well as semantic data integration techniques, to facilitate a common understanding of data. The design of dynamic protocols and middleware solutions can bridge the gap in realtime data synchronization to facilitate better CPS-SDT integration.
\par 

\textbf{Data and Modeling issues in Digital Twins:} Using diverse CPS data with SDTs adds data complexity and heterogeneity for training AI/ML and semantic models. In this regard, data curation, i.e. data labeling and distribution with sampling bias, have security implications. The integration of sensor source data and synthetic data should be used with techniques such as transfer learning and active learning to minimize false positives/false negatives resulting in more robust models \cite{arp2022and}. 
To address the constraints of limited labeled data, automated methods can be used effectively. This involves generating additional simulated data based on existing labeled data, and introducing variations, noise, or anomalies to increase the diversity of data.
The AI/ML models' robustness should be enhanced with emerging semantic and knowledge-based integration enabling context-aware models. These models encompass basic detection and prediction techniques to include increased refinement of machine learning using ML and machine reasoning technologies. 
Tailoring techniques to meet the specific requirements of cybersecurity within the digital twin operational environment can meaningfully contribute to the efficient handling of data and improved robust model development. \par
\textbf{Trustworthy Modeling for Digital Twins:}
The absence of transparency and interpretability (Blackbox) in current AI/ML algorithms when used in Digital Twins erodes the trust of security teams in the models employed to counteract cyberattacks. Foundational models utilized for prediction, detection, and analysis need to instil a measurable level of confidence among stakeholders regarding their relevance, reliability, and accuracy. However, most of these AI/ML models function as black boxes, offering limited insight into the reasoning behind their output. To address this shortcoming, Explainable AI (XAI) methods \cite{zhang2022explainable} should be developed and integrated into cyber defense, attack analysis, and cybersecurity incident response stages. SDTs, empowered by XAI methods, should offer unambiguous transparency, interpretability, and clarity of understanding at each phase of automated cybersecurity operations.  
\section{Conclusion and Future Research Directions}

In this vision paper, we present innovative concepts for crafting intelligent security DTs, poised to enhance cybersecurity operations by applying the SLSR design paradigm for securing systems. These security-focused Digital Twins aim to bolster trust and resilience within critical smart systems, encompassing IT/OT in critical infrastructure and Industry 4.0 operational domains.
Future systems cybersecurity is envisioned with virtual components and the SLSR design paradigm enabling reliable monitoring of CPS, fostering robust security twins model development across the system lifecycle. 
The future security-driven DTs shall have integrated support for diverse cybersecurity use cases employing AI/ML, semantic structures, and KGs with continuous automation and integration. By harnessing generative AI combined with KGs and modeling strategies, SDTs shall pave the way for innovative cybersecurity solutions for critical systems of the future. 
\par 
For future research, a number of issues are identified for exploration and development, as outlined in the research challenges sections. Data curation, management, and processing for SDT consumption require further application refinement and extensive research. The integration of generative AI for cybersecurity within the context of DT holds significant promise. However, the trustworthy and responsible design of these AI/ML models requires additional research before they can be seamlessly integrated into security Digital Twins. 


\section*{Acknowledgment}

The work has been supported by the Cyber Security Research Centre Limited whose activities are partially funded by the Australian Government’s Cooperative Research Centres Program.

\bibliography{vispaper.bib}

\begin{thebibliography}{10}

\bibitem{cpsics}
Y.~Wang, M.~C. Vuran, and S.~Goddard, ``Cyber-physical systems in industrial
  process control,'' {\em SIGBED Rev.}, vol.~5, jan 2008.

\bibitem{10034656}
M.~Jafari, A.~Kavousi-Fard, T.~Chen, and M.~Karimi, ``A review on digital twin
  technology in smart grid, transportation system and smart city: Challenges
  and future,'' {\em IEEE Access}, vol.~11, pp.~17471--17484, 2023.

\bibitem{ENISA2022}
T.~E. U.~A. for Cybersecurity~(ENISA), ``Enisa threat landscape 2022,'' tech.
  rep., ENISA, 2022.

\bibitem{app11219785}
J.~Hajda, R.~Jakuszewski, and S.~Ogonowski, ``Security challenges in industry
  4.0 plc systems,'' {\em Applied Sciences}, vol.~11, no.~21, 2021.

\bibitem{colonialattack}
C.~R. Service, ``Colonial pipeline: The darkside strikes,'' 2021.

\bibitem{solarwindattack}
S.~Gupta, ``Taxonomy of the attack on solarwinds and its supply chain,'' 2020.

\bibitem{Trittonattack}
M.~Giles, ``Triton is the world’s most murderous malware, and it’s
  spreading,'' 2019.

\bibitem{9763485}
W.~Duo, M.~Zhou, and A.~Abusorrah, ``A survey of cyber attacks on cyber
  physical systems: Recent advances and challenges,'' {\em IEEE/CAA Journal of
  Automatica Sinica}, vol.~9, no.~5, pp.~784--800, 2022.

\bibitem{LI20218176}
Y.~Li and Q.~Liu, ``A comprehensive review study of cyber-attacks and cyber
  security; emerging trends and recent developments,'' {\em Energy Reports},
  vol.~7, pp.~8176--8186, 2021.

\bibitem{IEC62443standard}
G.~S.~A. ISA, ``Aan overview of isa/iec 62443 standards security of industrial
  automation and control systems,'' tech. rep., Security of Industrial
  Automationand Control Systems, 2023.

\bibitem{nistframework}
{National Institute of Standards and Technology}, ``{The NIST Cybersecurity
  Framework 2.0 },'' 2023.

\bibitem{DTclassic}
B.~R. Barricelli, E.~Casiraghi, and D.~Fogli, ``A survey on digital twin:
  Definitions, characteristics, applications, and design implications,'' {\em
  IEEE Access}, vol.~7, pp.~167653--167671, 2019.

\bibitem{schluse2016simulation}
M.~Schluse and J.~Rossmann, ``From simulation to experimentable digital twins:
  Simulation-based development and operation of complex technical systems,'' in
  {\em 2016 IEEE international symposium on systems engineering (ISSE)},
  pp.~1--6, IEEE, 2016.

\bibitem{pires2019digital}
F.~Pires, A.~Cachada, J.~Barbosa, A.~P. Moreira, and P.~Leit{\~a}o, ``Digital
  twin in industry 4.0: Technologies, applications and challenges,'' in {\em
  2019 IEEE 17th International Conference on Industrial Informatics (INDIN)},
  vol.~1, pp.~721--726, IEEE, 2019.

\bibitem{holmes2021digital}
D.~Holmes, M.~Papathanasaki, L.~Maglaras, M.~A. Ferrag, S.~Nepal, and
  H.~Janicke, ``Digital twins and cyber security--solution or challenge?,'' in
  {\em 2021 6th South-East Europe Design Automation, Computer Engineering,
  Computer Networks and Social Media Conference (SEEDA-CECNSM)}, pp.~1--8,
  IEEE, 2021.

\bibitem{Purduemodel}
T.~J. Williams, ``The purdue enterprise reference architecture,'' {\em
  Computers in Industry}, vol.~24, no.~2, pp.~141--158, 1994.

\bibitem{smartgrid2}
A.~Alhariry, S.~Brown, D.~Eshenbaugh, N.~Whitt, and A.~F. Browne, ``A survey of
  sensing methodologies in smart grids,'' in {\em SoutheastCon 2021}, pp.~1--5,
  2021.

\bibitem{smartgrid1}
M.~N. Nafees, N.~Saxena, A.~Cardenas, S.~Grijalva, and P.~Burnap, ``Smart grid
  cyber-physical situational awareness of complex operational technology
  attacks: A review,'' {\em ACM Computing Surveys}, vol.~55, no.~10, 2023.

\bibitem{PIVOTO2021176}
D.~G. Pivoto, L.~F. {de Almeida}, R.~{da Rosa Righi}, J.~J. Rodrigues, A.~B.
  Lugli, and A.~M. Alberti, ``Cyber-physical systems architectures for
  industrial internet of things applications in industry 4.0: A literature
  review,'' {\em Journal of Manufacturing Systems}, vol.~58, pp.~176--192,
  2021.

\bibitem{9211890}
C.~Scordino, I.~M. Savino, L.~Cuomo, L.~Miccio, A.~Tagliavini, M.~Bertogna, and
  M.~Solieri, ``Real-time virtualization for industrial automation,'' in {\em
  2020 25th IEEE International Conference on Emerging Technologies and Factory
  Automation (ETFA)}, vol.~1, pp.~353--360, 2020.

\bibitem{virtualisationIIOts}
M.~Cinque, D.~Cotroneo, L.~{De Simone}, and S.~Rosiello, ``Virtualizing
  mixed-criticality systems: A survey on industrial trends and issues,'' {\em
  Future Generation Computer Systems}, vol.~129, pp.~315--330, 2022.

\bibitem{DTSOAR}
P.~Empl, D.~Schlette, D.~Zupfer, and G.~Pernul, ``Soar4iot: Securing iot assets
  with digital twins,'' in {\em Proceedings of the 17th International
  Conference on Availability, Reliability and Security}, ARES '22, (New York,
  NY, USA), Association for Computing Machinery, 2022.

\bibitem{8260283}
R.~Khan, K.~McLaughlin, D.~Laverty, and S.~Sezer, ``Stride-based threat
  modeling for cyber-physical systems,'' in {\em 2017 IEEE PES Innovative Smart
  Grid Technologies Conference Europe (ISGT-Europe)}, pp.~1--6, 2017.

\bibitem{Kaur2020}
M.~J. Kaur, V.~P. Mishra, and P.~Maheshwari, {\em The Convergence of Digital
  Twin, IoT, and Machine Learning: Transforming Data into Action}, pp.~3--17.
\newblock Cham: Springer International Publishing, 2020.

\bibitem{9453553}
W.~Tärneberg, P.~Skarin, C.~Gehrmann, and M.~Kihl, ``Prototyping intrusion
  detection in an industrial cloud-native digital twin,'' in {\em 2021 22nd
  IEEE International Conference on Industrial Technology (ICIT)}, vol.~1,
  pp.~749--755, 2021.

\bibitem{info14020095}
P.~Empl and G.~Pernul, ``Digital-twin-based security analytics for the internet
  of things,'' {\em Information}, vol.~14, no.~2, 2023.

\bibitem{9527938}
A.~Patel, T.~Schenk, S.~Knorn, H.~Patzlaff, D.~Obradovic, and A.~B. Halblaub,
  ``Real-time, simulation-based identification of cyber-security attacks of
  industrial plants,'' in {\em 2021 IEEE International Conference on Cyber
  Security and Resilience (CSR)}, pp.~267--272, 2021.

\bibitem{IDS-staic-DT}
M.~Eckhart and A.~Ekelhart, ``Towards security-aware virtual environments for
  digital twins,'' in {\em Proceedings of the 4th ACM Workshop on
  Cyber-Physical System Security}, CPSS '18, (New York, NY, USA), p.~61–72,
  Association for Computing Machinery, 2018.

\bibitem{9438560}
Q.~Xu, S.~Ali, and T.~Yue, ``Digital twin-based anomaly detection in
  cyber-physical systems,'' in {\em 2021 14th IEEE Conference on Software
  Testing, Verification and Validation (ICST)}, pp.~205--216, 2021.

\bibitem{Varghese_2022}
S.~A. Varghese, A.~D. Ghadim, A.~Balador, Z.~Alimadadi, and P.~Papadimitratos,
  ``Digital twin-based intrusion detection for industrial control systems,'' in
  {\em 2022 {IEEE} International Conference on Pervasive Computing and
  Communications Workshops and other Affiliated Events ({PerCom} Workshops)},
  {IEEE}, mar 2022.

\bibitem{DTSOC}
M.~Vielberth, M.~Glas, M.~Dietz, S.~Karagiannis, E.~Magkos, and G.~Pernul, ``A
  digital twin-based cyber range for soc analysts,'' in {\em Data and
  Applications Security and Privacy XXXV} (K.~Barker and K.~Ghazinour, eds.),
  (Cham), pp.~293--311, Springer International Publishing, 2021.

\bibitem{10049398}
E.~C. Balta, M.~Pease, J.~Moyne, K.~Barton, and D.~M. Tilbury, ``Digital
  twin-based cyber-attack detection framework for cyber-physical manufacturing
  systems,'' {\em IEEE Transactions on Automation Science and Engineering},
  pp.~1--18, 2023.

\bibitem{austin2020architecting}
M.~Austin, P.~Delgoshaei, M.~Coelho, and M.~Heidarinejad, ``Architecting smart
  city digital twins: Combined semantic model and machine learning approach,''
  {\em Journal of Management in Engineering}, vol.~36, no.~4, p.~04020026,
  2020.

\bibitem{MLcybersec}
G.~Apruzzese, P.~Laskov, E.~Montes~de Oca, W.~Mallouli, L.~Brdalo~Rapa, A.~V.
  Grammatopoulos, and F.~Di~Franco, ``The role of machine learning in
  cybersecurity,'' {\em Digital Threats}, vol.~4, mar 2023.

\bibitem{kucharavy2023fundamentals}
A.~Kucharavy, Z.~Schillaci, L.~Maréchal, M.~Würsch, L.~Dolamic,
  R.~Sabonnadiere, D.~P. David, A.~Mermoud, and V.~Lenders, ``Fundamentals of
  generative large language models and perspectives in cyber-defense,'' 2023.

\bibitem{openai2023}
{OpenAI}, ``{ChatGPT: A Language Model for Conversational AI},'' tech. rep.,
  {OpenAI}, 2023.

\bibitem{semantickGsDT}
X.~Zheng, J.~Lu, and D.~Kiritsis, ``The emergence of cognitive digital twin:
  vision, challenges and opportunities,'' {\em International Journal of
  Production Research}, vol.~60, no.~24, pp.~7610--7632, 2022.

\bibitem{mcdonald2010modeling}
M.~McDonald, J.~Mulder, B.~Richardson, R.~Cassidy, A.~Chavez, N.~Pattengale,
  G.~Pollock, J.~Urrea, M.~Schwartz, W.~Atkins, {\em et~al.}, ``Modeling and
  simulation for cyber-physical system security research, development and
  applications,'' {\em Sandia National Laboratories, Tech. Rep. Sandia Report
  SAND2010-0568}, 2010.

\bibitem{YAMIN2020101636}
M.~M. Yamin, B.~Katt, and V.~Gkioulos, ``Cyber ranges and security testbeds:
  Scenarios, functions, tools and architecture,'' {\em Computers \& Security},
  vol.~88, p.~101636, 2020.

\bibitem{8842802}
T.~Debatty and W.~Mees, ``Building a cyber range for training cyberdefense
  situation awareness,'' in {\em 2019 International Conference on Military
  Communications and Information Systems (ICMCIS)}, pp.~1--6, 2019.

\bibitem{blockchainSDT}
S.~Suhail, R.~Hussain, R.~Jurdak, and C.~S. Hong, ``Trustworthy digital twins
  in the industrial internet of things with blockchain,'' {\em IEEE Internet
  Computing}, vol.~26, no.~3, pp.~58--67, 2022.

\bibitem{arp2022and}
D.~Arp, E.~Quiring, F.~Pendlebury, A.~Warnecke, F.~Pierazzi, C.~Wressnegger,
  L.~Cavallaro, and K.~Rieck, ``Dos and don'ts of machine learning in computer
  security,'' in {\em 31st USENIX Security Symposium (USENIX Security 22)},
  pp.~3971--3988, 2022.

\bibitem{zhang2022explainable}
Z.~Zhang, H.~Al~Hamadi, E.~Damiani, C.~Y. Yeun, and F.~Taher, ``Explainable
  artificial intelligence applications in cyber security: State-of-the-art in
  research,'' {\em IEEE Access}, 2022.

\end{thebibliography}
\bibliographystyle{ieeetr}
\end{document}